\documentclass[12pt,a4paper]{article}
\usepackage{hyperref}
\usepackage{siunitx}
\usepackage{mathptmx}
\usepackage{graphicx} 
\usepackage{soul} 
\usepackage{float}

\usepackage{amsmath}
\usepackage{amsfonts}
\usepackage{amssymb}
\usepackage[a4paper, left=25mm, right=25mm, top=25mm, bottom=20mm]{geometry}

\begin{document}

\begin{titlepage}

{\flushright{
        \begin{minipage}{2.5cm}
        \end{minipage}        }

}

\begin{center}
{\LARGE\bf
Optimization of Undulator Parameters for 125 GeV Drive Beam\footnote{Talk presented at the International Workshop on Future Linear Colliders (LCWS2018), Arlington, Texas, 22-26 October 2018.}}
\vskip 1.0cm
{\large Manuel Formela$^1$\footnote{ manuel.formela@desy.de},
        Gudrid Moortgat-Pick$^{1,2}$, Sabine Riemann$^3$, Andriy Ushakov$^{1}$.}
        \vspace*{8mm}\\
{\sl\small ${}^1$ University of Hamburg, Luruper Chaussee 149, D-22761 Hamburg, Germany \\
${}^2$ Deutsches Elektronen-Synchrotron (DESY), Notkestraße 85, D-22607 Hamburg, Germany\\
${}^3$ Deutsches Elektronen-Synchrotron (DESY), Platanenallee 6, D-15738 Zeuthen, Germany}

\end{center}

\begin{abstract}
In the baseline design of the International Linear Collider (ILC) an undulator-based source is foreseen for the positron source in order to match the physics requirements. 
The baseline parameters are optimized for the ILC at 
$\sqrt{s}=500$~GeV, that means an electron drive beam  of $250$~GeV.  Precision measurements in the Higgs sector, however, require measurements at 
$\sqrt{s}=250$~GeV, i.\ e.\ running with the electron drive beam only at 125~GeV which imposes a challenge for achieving a high yield. 
Therefore the baseline undulator parameters have to be optimized as much as it is possible within their technical performances.

In this bachelor thesis we therefore present
a theoretical study on the radiation spectra of a helical undulator, based on the equation for the radiated synchrotron energy spectral density per solid angle per electron in the relativistic, far field and point-like charge approximation. From this starting point the following undulator properties are examined: 
the deposited power in the undulator vessel, which can disrupt the functionality of the undulator‘s magnets, and 
the protective property of a mask on such disturbances and the number of positrons produced by the synchrotron radiation in a Ti-6Al-4V target. Those quantities were evaluated for various values for parameters as undulator period, undulator length and magnetic flux in order to find optimal baseline parameter sets for $\sqrt{s}=250$~GeV.
\end{abstract}

\end{titlepage}


%
\section{Introduction}
The International Linear Collider (ILC) as well as further future high-energy colliders as CLIC, for instance, have to provide 
polarized beams at high intensity as well as at high energy. Challenging is the production of the high-intense positron beam.
The ILC uses an undulator-based positron source in the baseline design~\cite{Adolphsen:2013kya} that even produces a polarized positron beam. In this way, i.e. offering high intense and polarized electron and positron beams, the physics potential of the ILC is optimized and well prepared for high precision physics as well as new 
discoveries~\cite{Moortgat-Picka:2015yla,AguilarSaavedra:2001rg}. Currently an initial energy of $\sqrt{s}=250$~GeV is discussed~\cite{Fujii:2018mli}, where the undulator scheme
can be applied as well~\cite{Ushakov:2018wlt}.

However, one should note, that the precision requirements can only be fulfilled if polarized positrons are available already at $\sqrt{s}=250$~GeV!
Otherwise the systematic uncertainties get too large, see \cite{Sabine, Robert-Thesis,Karl:2017xra,updateTDR}. Applying simultaneously-polarized beams allows to
be competitive to $e^+e^-$ circular design-proposals that offer an order of magnitude higher luminosity but cannot provide polarized beams.
Therefore it is of utmost importance to optimize the chosen baseline undulator-parameter set already for this first stage energy.

In the following we will have a look at different aspects on calculating  the corresponding properties of helical undulators. 
First we discuss the analytical formulas of undulator radiation and the resulting produced positron number for a given undulator set-up.
In section~\ref{sect:rdr_und}, we list the basic formulae used for describing undulator radiation, in section~\ref{sect:und}, we 
compare our undulator results with the values in the literature for the BCD- and RDR-design of the ILC. In section~\ref{sect:opt} we perform
several undulator parameter scans in order to further optimize the baseline design --that was foreseen for $\sqrt{s}=500$~GeV-- for the currently discussed energy stage of 
$\sqrt{s}=250$~GeV.

\section{Undulator-based positron source scheme: fundamentals \label{sect:rdr_und}}

\begin{figure}[H]
\fbox{
		\includegraphics[scale=0.5,page=3]{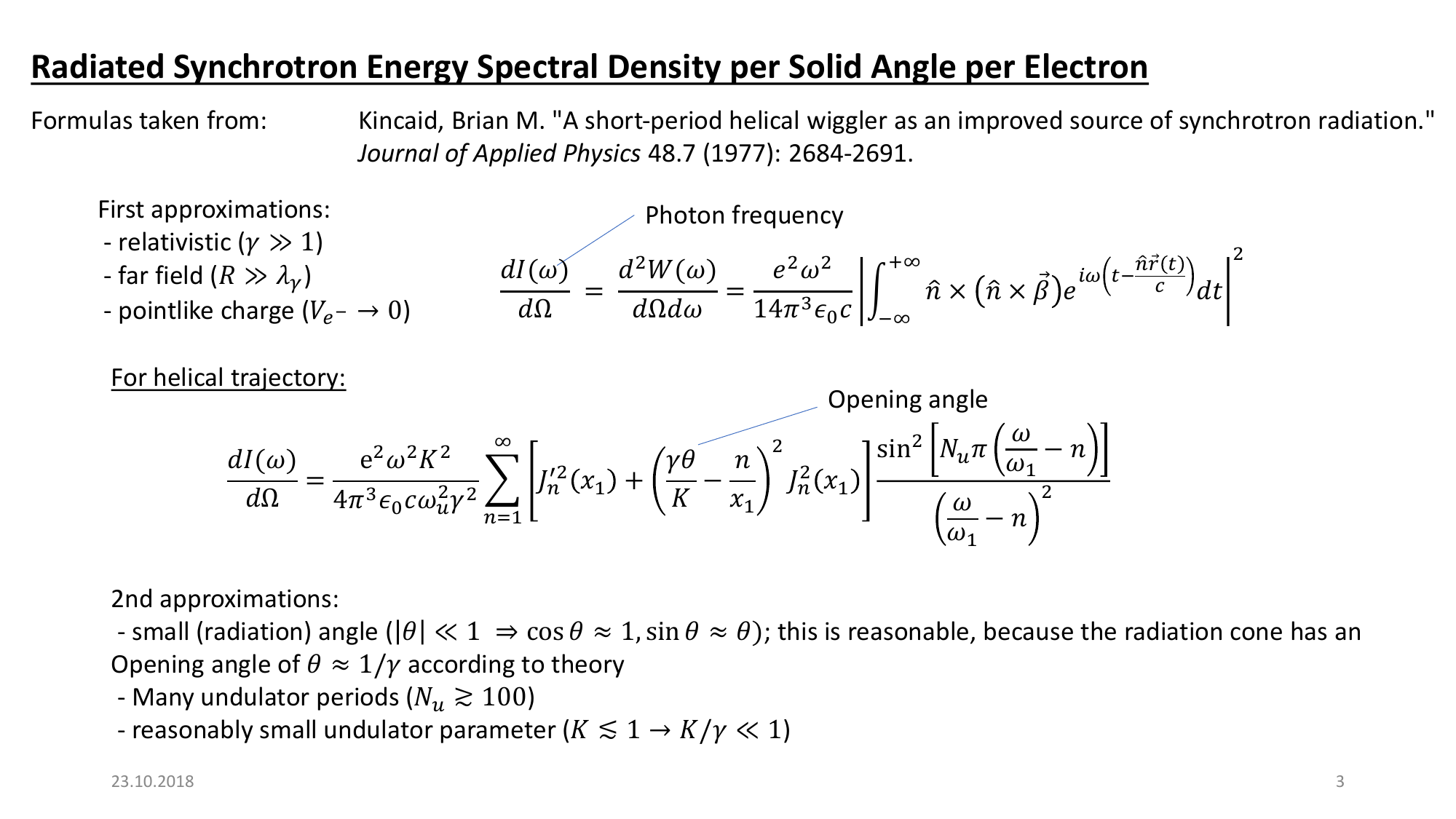}
		}
\vspace{-.5cm}	
\caption{ Analytical expression to calculate the radiated spectral energy density caused 
by one electron when travelling through a short-period helical undulator with at least 100 periods and with a magnetic field  resulting in the 
undulator parameter $K\le 1$~\cite{Kincaid:1977fg}.\label{fig_1}}
\end{figure}
\noindent
The first formula in Fig.~\ref{fig_1} shows the spectral energy density per solid angle $\Omega$ of synchrotron radiation produced by one electron 
$\frac{dI(\omega)}{d\Omega} = \frac{d^2W(\omega)}{d\Omega d\omega}$ (in the far-field approximation, $R >> \lambda_\gamma$, for pointlike, $V_{e^-} \rightarrow 0$,  relativistic, $\gamma << 1$, electrons) \cite{Kincaid:1977fg}. The given formula describes the synchrotron radiation with the
photon frequency $\omega$ in general, valid for any trajectory of the electric charge.   

The second equation in Fig.~\ref{fig_1}  expresses the spectral energy density applied for a specific charge trajectory, i.\ e.\ for  a helical trajectory of the undulator electron. 
It contains the $n$-th order Bessel function of the first kind $J_n$ and its first derivative $J_n^\prime$. This equation holds for small radiation angles ($|\theta| <<1$), many undulator periods ($N_u \gtrsim 100$) and for reasonably small undulator parameter ($K \lesssim 1$)~\cite{Kincaid:1977fg}.

\begin{figure}[H]
	\fbox{
		\includegraphics[scale=0.5,page=4]{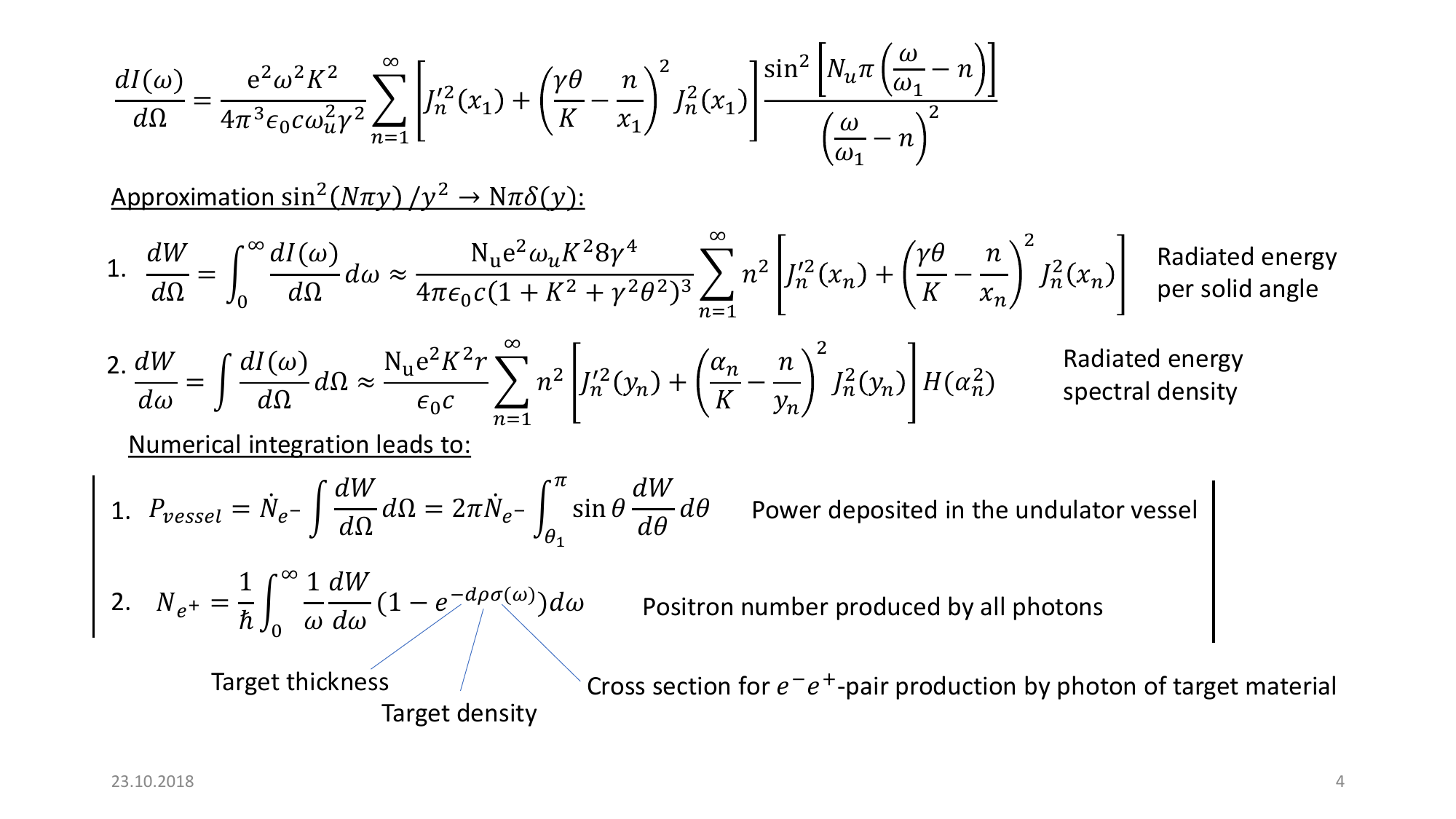}}
\caption{ Analytical expressions for deriving 1) the total power that is deposited in the undulator vessel and 2) the total number of produced positrons by the 
undulator radiation~\cite{Kincaid:1977fg}.    
\label{fig_2}}
\end{figure}
\noindent
By integrating the previous formula over all photon frequencies $\omega$, one obtains the radiated energy per solid angle per electric charge $\frac{dW}{d\Omega}$,
see Fig.~\ref{fig_2}. For the integration, the fraction with the squared sinus function in the numerator was approximated by the Dirac delta distribution ($\sin^2(N\pi y)/y^2 \rightarrow N\pi \delta (y)$), which simplifies the integration. \\
Integrating over the whole solid angle leads to the radiated energy spectral density $\frac{dW}{d\omega}$. Here the step function $H$ occurs. 

Integrating numerically the radiated energy per solid angle $\frac{dW}{d\Omega}$ over the solid angle, which covers the undulator vessel, results in the power deposited in the undulator vessel $P_{vessel}$, when multiplied by the electron rate $\dot N_{e^-}$. Since this integral has no analytical solution, we solve it numerically.

Analogously, we get 
the produced positron number $N_{e^+}$ by dividing 
the radiated energy spectral density $\frac{dW}{d\omega}$ by the photon energy $\hbar \omega$ and by multiplying with the 
probability for electron positron pair production $(1 - e^{-d\rho \sigma (\omega)})$ integrated over the photon frequency $\omega$. 
The probability function is dependent on the cross section for electron positron pair production of a specific target 
material $\sigma (\omega)$, the target density $\rho$ and its thickness $d$. 

\section{Undulator scheme used in the RDR \label{sect:und}}
\begin{figure}[H]
	\fbox{
		\includegraphics[scale=0.5,page=5]{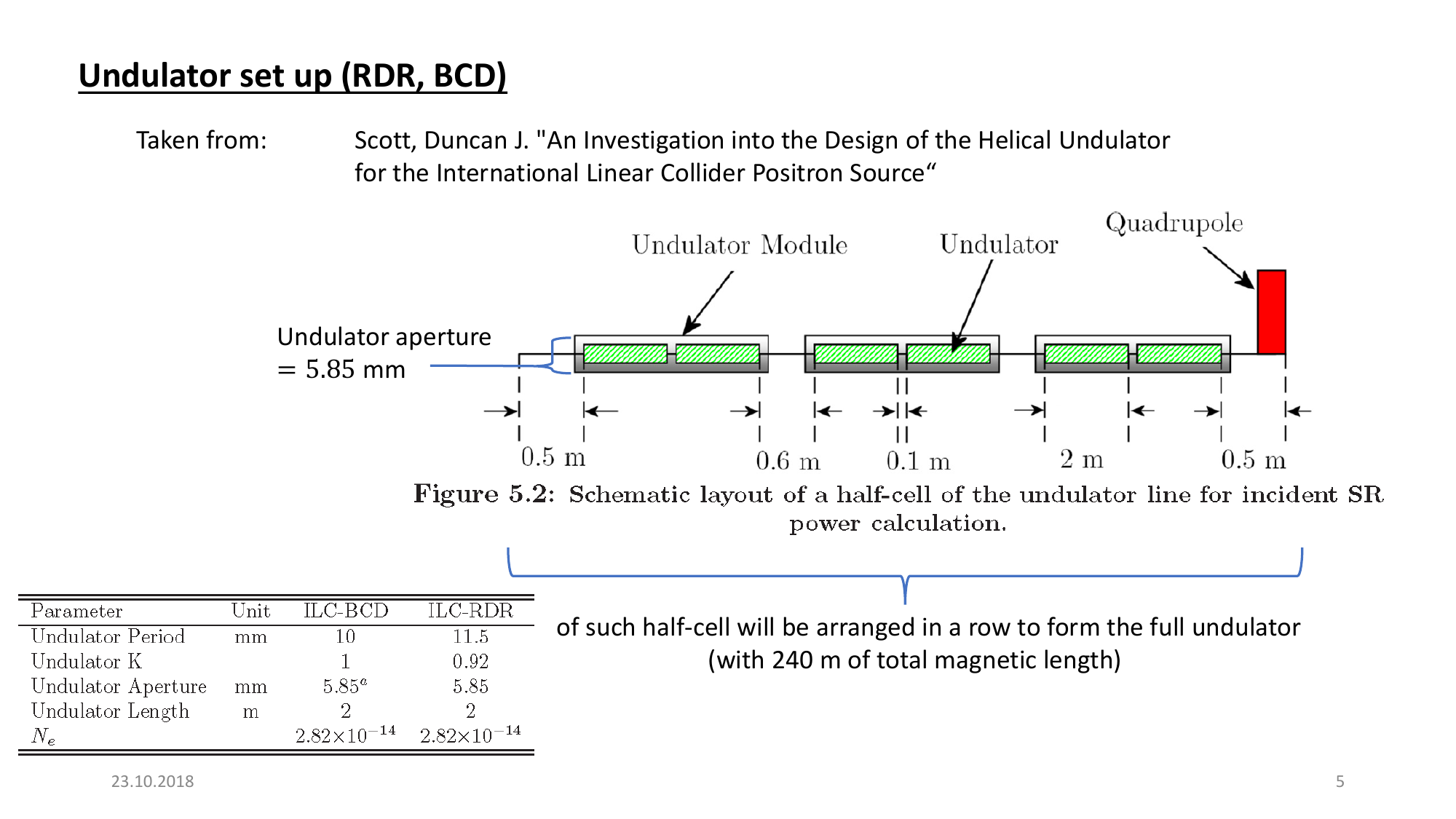}}
\caption{Parameters and geometrical set-up for the helical undulator design chosen for the Baseline Conceptual Design (BCD) and the 
Reference Design Report (RDR) of the ILC \cite{BCD, ilc-rdr}. \label{fig_3}}
\end{figure}
\noindent
The undulator set-up and the undulator parameters  shown in Fig.~\ref{fig_3} have been used for the ILC Baseline Conceptual Design (BCD)~\cite{BCD} and the 
Reference Design Report (RDR)~\cite{ilc-rdr}, more details are given in~\cite{duncan-thesis}. This setup has been used  to compare 
the detailed studies in~\cite{duncan-thesis} with our results, derived with the 
analytical formulae given in Figs.\ref{fig_1} and \ref{fig_2}, for
the power deposited in the undulator vessel $P_{vessel}$ and for the produced positron number $N_{e^+}$, where each set-up contains  20 half-cells, which
in turn consist of 3 undulator modules, respectively. Each undulator module is made up of 2 undulator magnets of $\SI{2}{\m}$ length separated by $\SI{0.1}{\m}$, 
see Fig.~\ref{fig_3}. The distance between neighbouring modules is $\SI{0.6}{\m}$ and finally each half-cell starts and concludes with an empty segments of $\SI{0.5}{\m}$ length. So the total length of a half-cell is $\SI{14.5}{\m}$, which adds up for all half-cells to the total undulator length of $\SI{290}{\m}$ with a total magnetic length of $\SI{240}{\m}$.


\subsection{Reproducing values for $P_{vessel}$ for the RDR set-up}

\begin{figure}[H]
	\fbox{
		\includegraphics[scale=0.5,page=6]{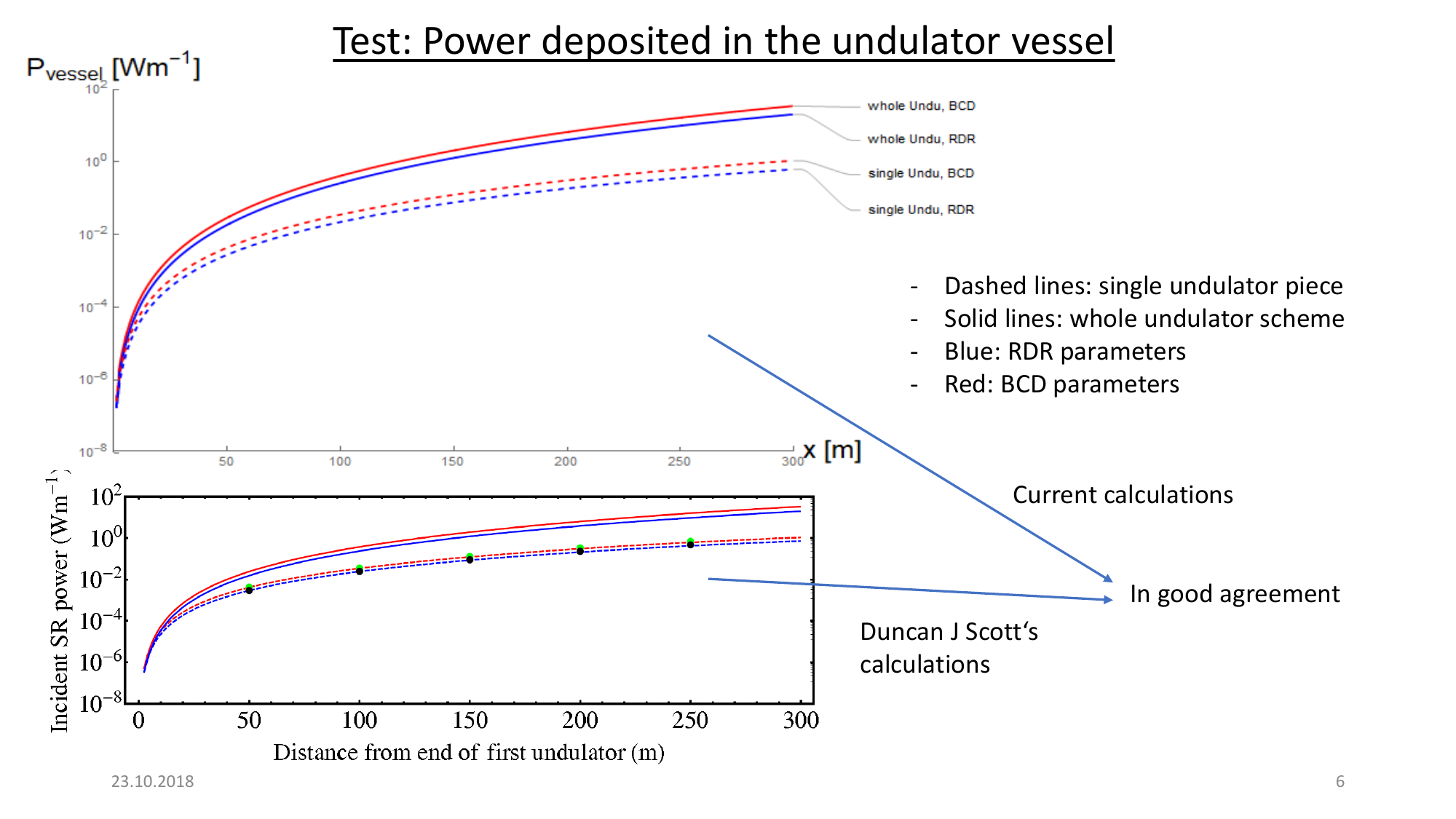}}
\caption{Comparison of deposited power in the undulator vessel derived in~\cite{duncan-thesis} and in our study~\cite{bachelor}.\label{fig_4}}
\end{figure}
\noindent
In Fig.~\ref{fig_4}, we compare our results ~\cite{bachelor} 
for the deposited power in the undulator vessel in the BCD- and RDR-undulator set-ups for the ILC~\cite{BCD, ilc-rdr} with
the original studies presented in \cite{duncan-thesis}:
both graphs show logarithmic plots of the deposited power in the undulator vessel per meter versus $x=$the distance from the exit of the first $\SI{2}{\m}$ long undulator magnet. In each of the two graphs there are four curves in red(BCD) and blue (RDR): dashed lines signify that only one $\SI{2}{\m}$ long undulator magnet was considered, while solid lines represent calculations for the whole undulator set up. 

The upper plot~\cite{bachelor} and lower plot~\cite{duncan-thesis} show a good agreement of both results, not only graphically but also after more rigorous comparing. 
One important result from these plots confirms that the whole undulator radiation is expected to deposit a total power in the vessel of values above $10$~W/m for BCD and RDR parameters.

\subsection{Implementing of undulator mask}
\begin{figure}[H]
	\fbox{
		\includegraphics[scale=0.5,page=7]{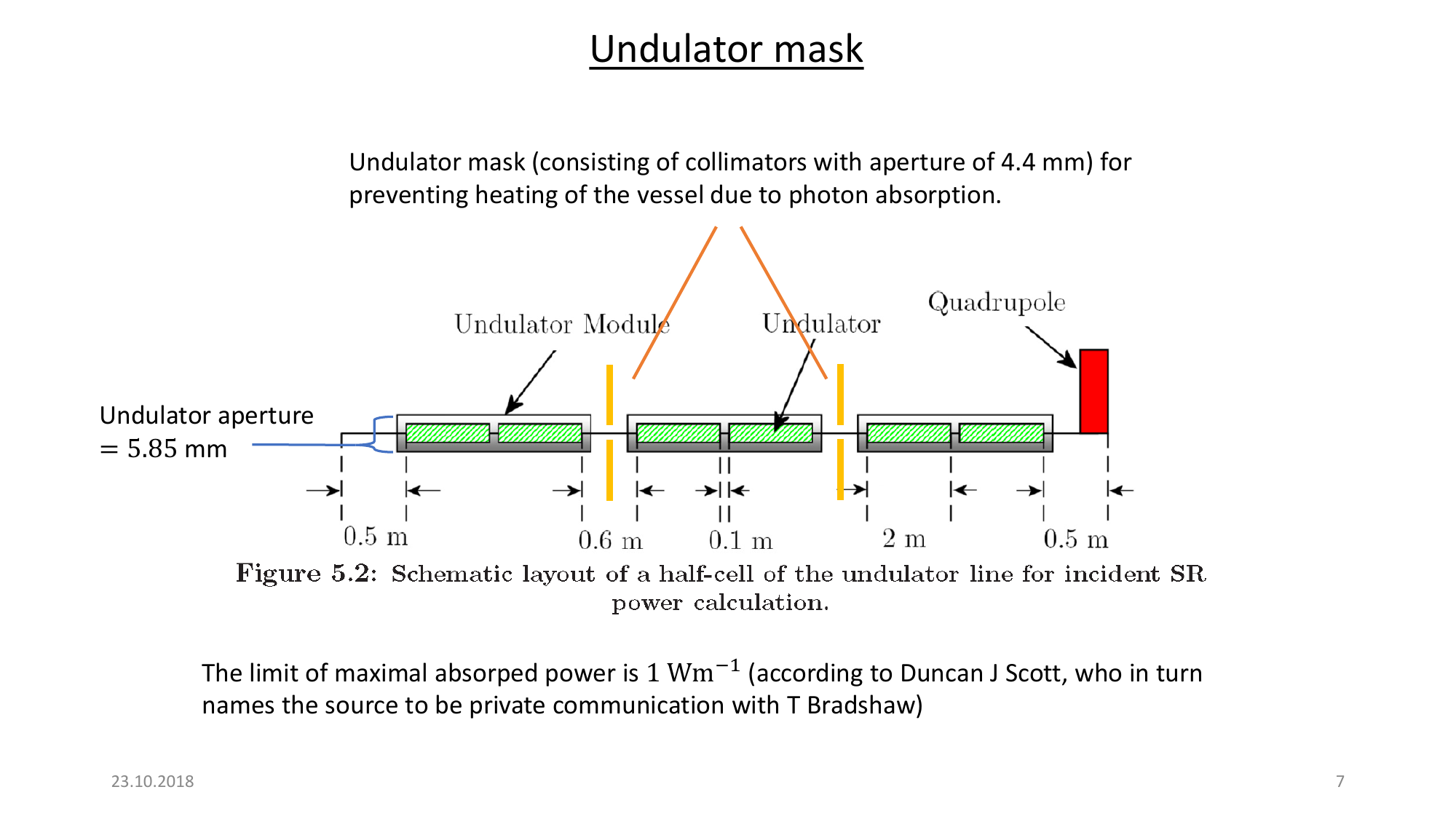}}
		\caption{Used scheme for implementing an undulator mask to protect the vessel from high deposited powers~\cite{duncan-thesis}.\label{fig_5}}
\end{figure}

Since the deposited power $P_{vessel}$ reaches, however, some tens W/m at the end of the undulator, a mask must be implemented.
Otherwise powers of $\SI{1}{\W \per \m}$ and above would disrupt the functionality of the magnetic coils and the vacuum. Such a limit was recommended by \cite{Tom}.
Fig.~\ref{fig_5} shows the geometrical layout for the implementation of the mask.
The mask consists of some material reaching into the beam axis and therefore protecting the following vessel part from the synchrotron radiation. 
In this set-up, we have two undulator masks in each half-cell, located between the modules.

We calculated again the expected power deposited in the undulator vessel $P_{vessel}$ with the inclusion of the mask, see Fig.~\ref{fig_6}:
\begin{figure}[H]
	\fbox{
		\includegraphics[scale=0.5,page=8]{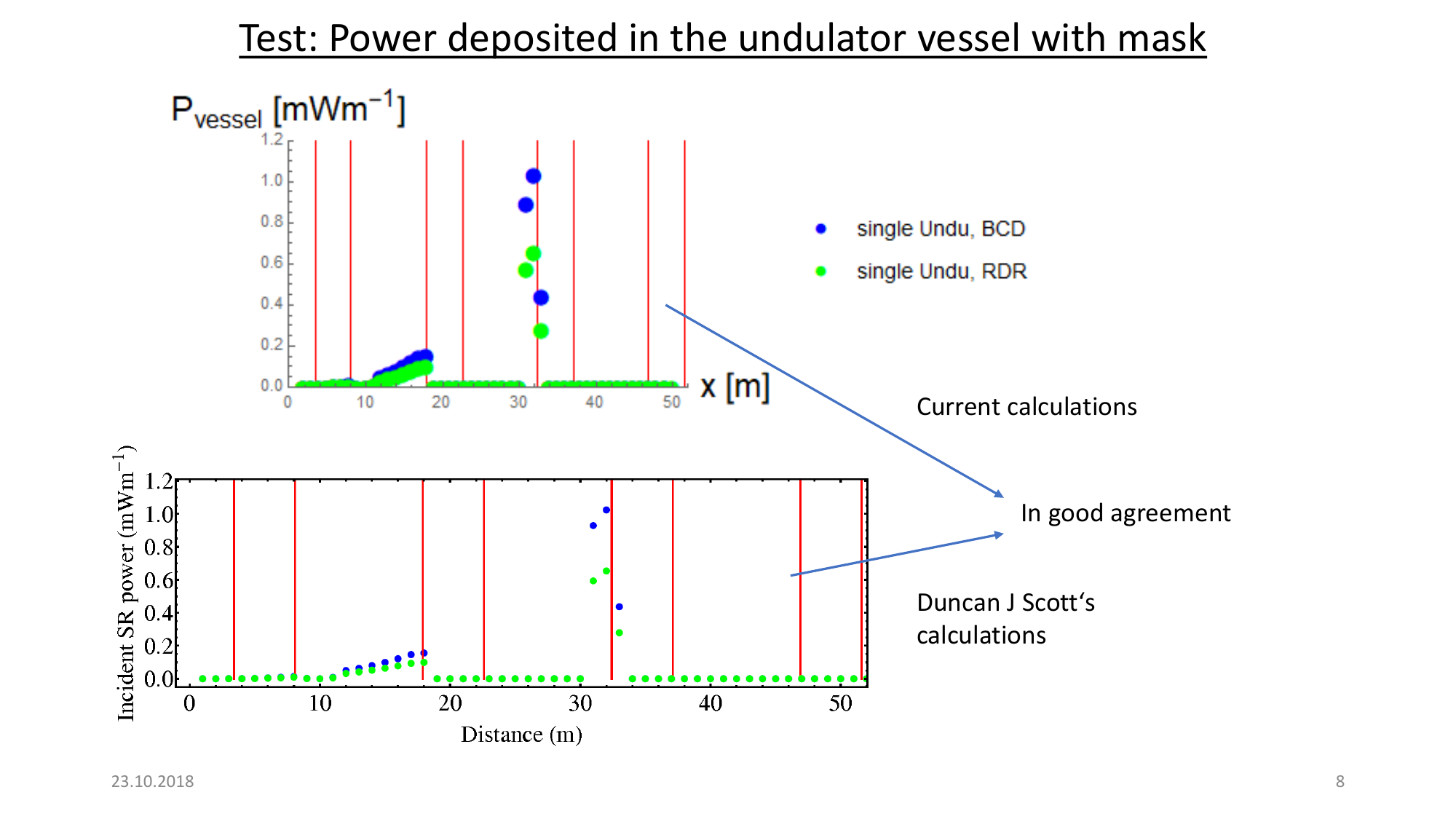}}
\caption{Deposited power in the undulator vessel when including an undulator mask (upper \cite{bachelor} and lower \cite{duncan-thesis} panel) . \label{fig_6}}
\end{figure}
\noindent

As in the case discussed before,  the two graphics (upper \cite{bachelor} and lower \cite{duncan-thesis} panel) in Fig.~\ref{fig_6} 
 are again in a good agreement and show the power deposited in the undulator versus $x=$the distance from the exit of the first $\SI{2}{\m}$ long undulator piece. 
Each graph includes two sets of data points, both calculated for one single $\SI{2}{\m}$ long undulator piece, but the blue set is based on the BCD parameters and the green one on the RDR parameters. The red vertical markers show the position of the mask. Obviously the mask prevents the vessel successfully from high power deposition.

After examining the effects of the mask on the deposited power for only one single $\SI{2}{\m}$-long undulator piece, 
one has also to calculate the effects of such a mass scheme for the whole undulator set-up. 
\begin{figure}[H]
	\fbox{
		\includegraphics[scale=0.5,page=9]{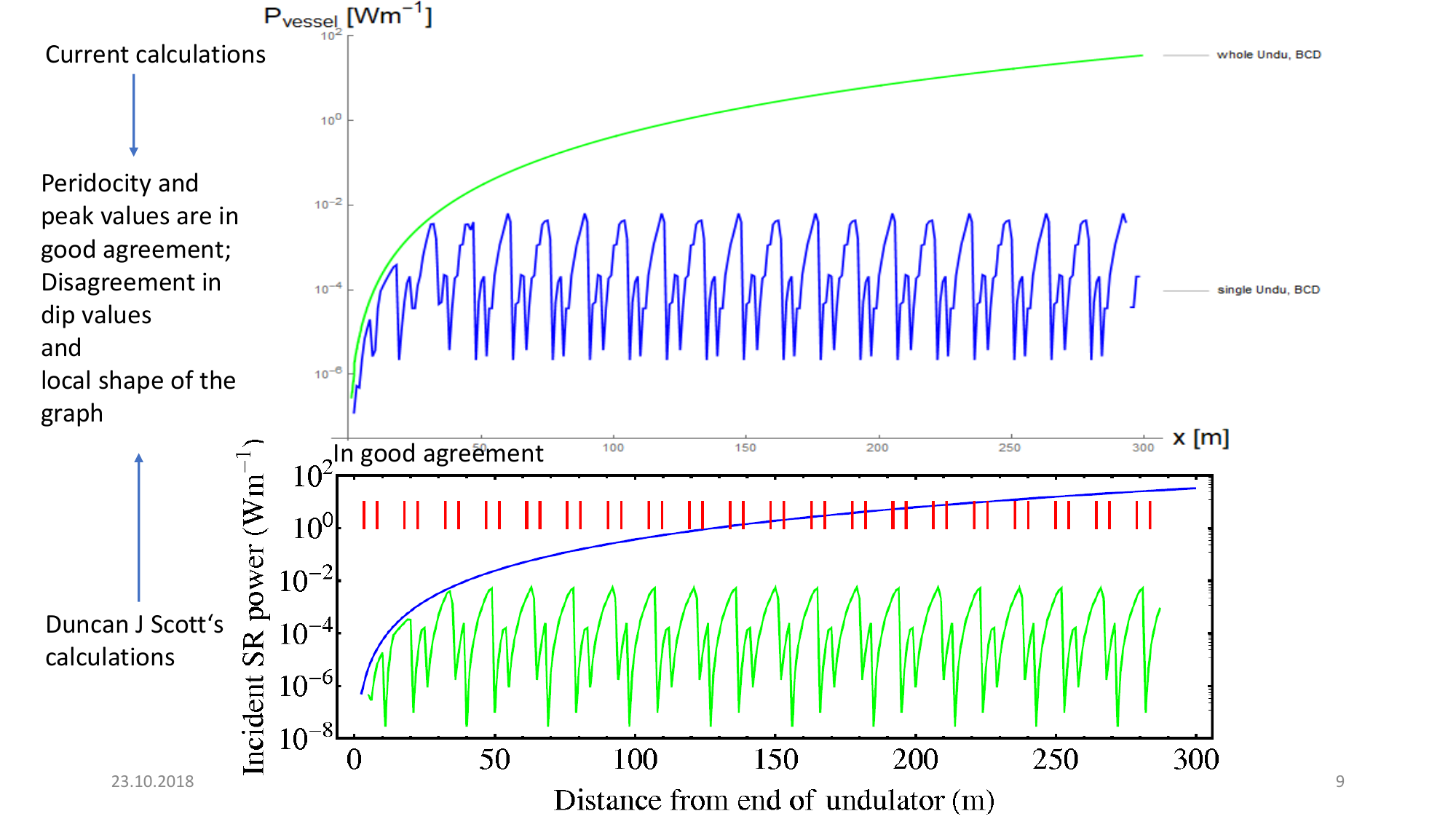}}
\caption{Total power deposition in the undulator vessel after including the mask scheme for the whole undulator in the BCD
(upper plot \cite{bachelor}, lower plot~\cite{duncan-thesis}). \label{fig_7}}
\end{figure}
In Fig.~\ref{fig_7} the total power deposited in the vessel per meter is plotted versus $x=$the distance from the exit of the first $\SI{2}{\m}$-long undulator piece
(upper panel \cite{bachelor}, lower panel\cite{duncan-thesis}). 
In the upper plot~\cite{bachelor} (lower plot ~\cite{duncan-thesis}), the blue (green) curve denotes the case when including the mask scheme. In both plots
also the calculation of the deposited power without the mask has been added for comparison, i.\ e.\ the green line in the upper plot ~\cite{bachelor} and the blue line in 
the lower plot~\cite{duncan-thesis}. Both curves are made for the BCD parameters~\cite{BCD}. 

In both graphics, it can be seen that the mask limits the maximal power deposited per meter of vessel below one hundredth of W/m, and therefore succeeds in meeting the previously discussed limit for the deposited energy per meter of $1$ W/m. 

Comparing both results for the case with the mask, show that there is a good agreement concerning the 
periodicity and peak values, but there are differences in dip values and the local shape of the graph. The cause for the disagreements might be due to different plot resolutions. 
Note, however, that the crucial important quantity is the peak value, which has to be below the upper limit condition.
The peak value fulfills this condition and is similar in both independent calculations.

\begin{figure}[H]
	\fbox{
		\includegraphics[scale=0.5,page=10]{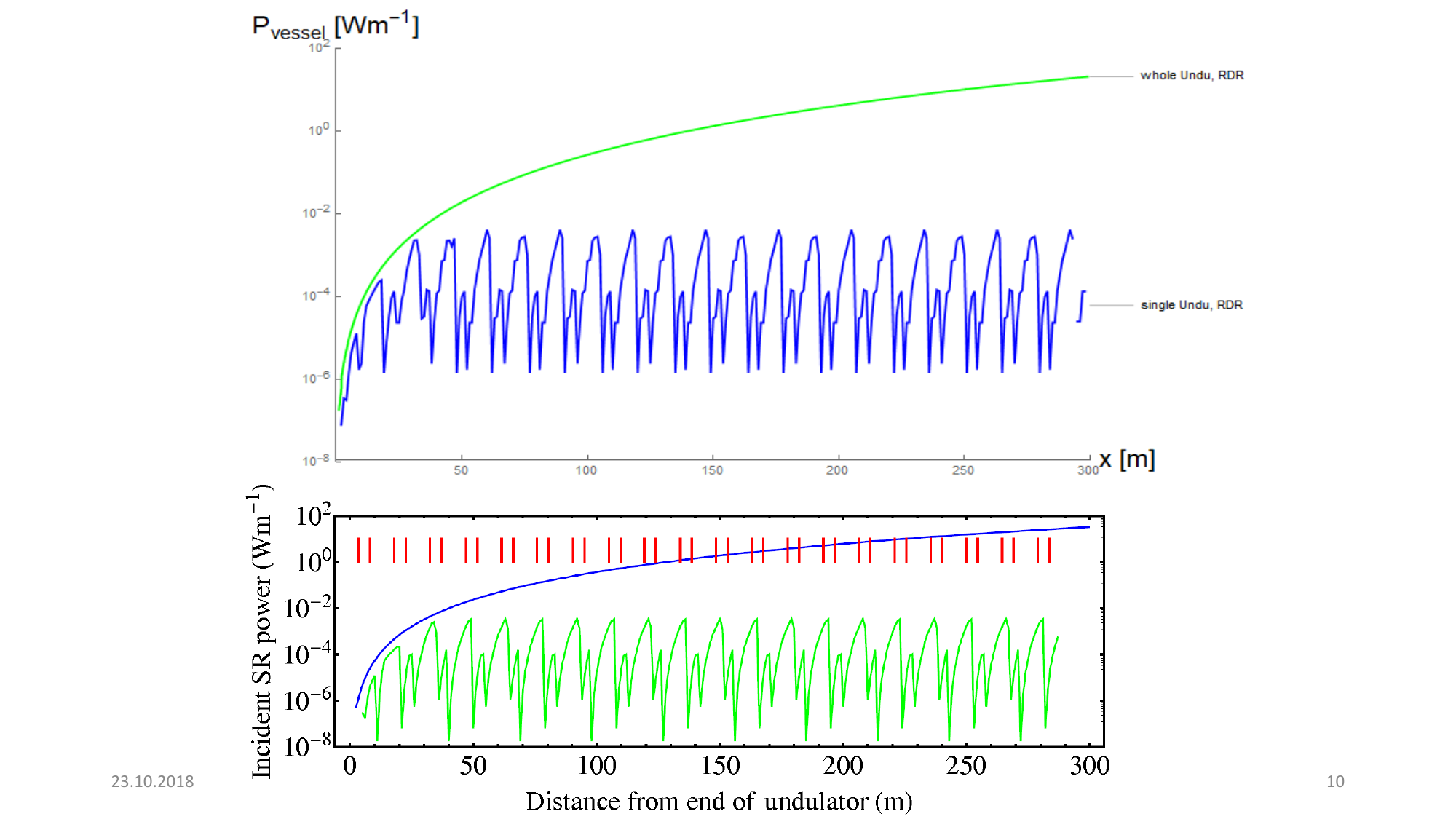}}
\caption{Total power deposition in the undulator vessel after inclusion of the mask for the whole undulator in the RDR (upper plot \cite{bachelor}, lower plot~\cite{duncan-thesis}). \label{fig_8}}
\end{figure}

In Fig.~\ref{fig_8} the same calculations were applied for the undulator including the mask scheme but for the RDR parameters~\cite{ilc-rdr}. 
The same conclusion as before can be drawn: the mask
reduces successfully the deposited power below the agreed limit of $1$ W/m~\cite{Tom}.
\noindent

\section{Steps towards optimizing the given undulator set-up for $\sqrt{s}=250$~GeV \label{sect:opt}}

\subsection{Calculation of $N_{e^+}$ for various parameter values for $K$, $\lambda$, $l_U$, $N_{hcell}$}
\begin{figure}
	\fbox{
		\includegraphics[scale=0.5,page=11]{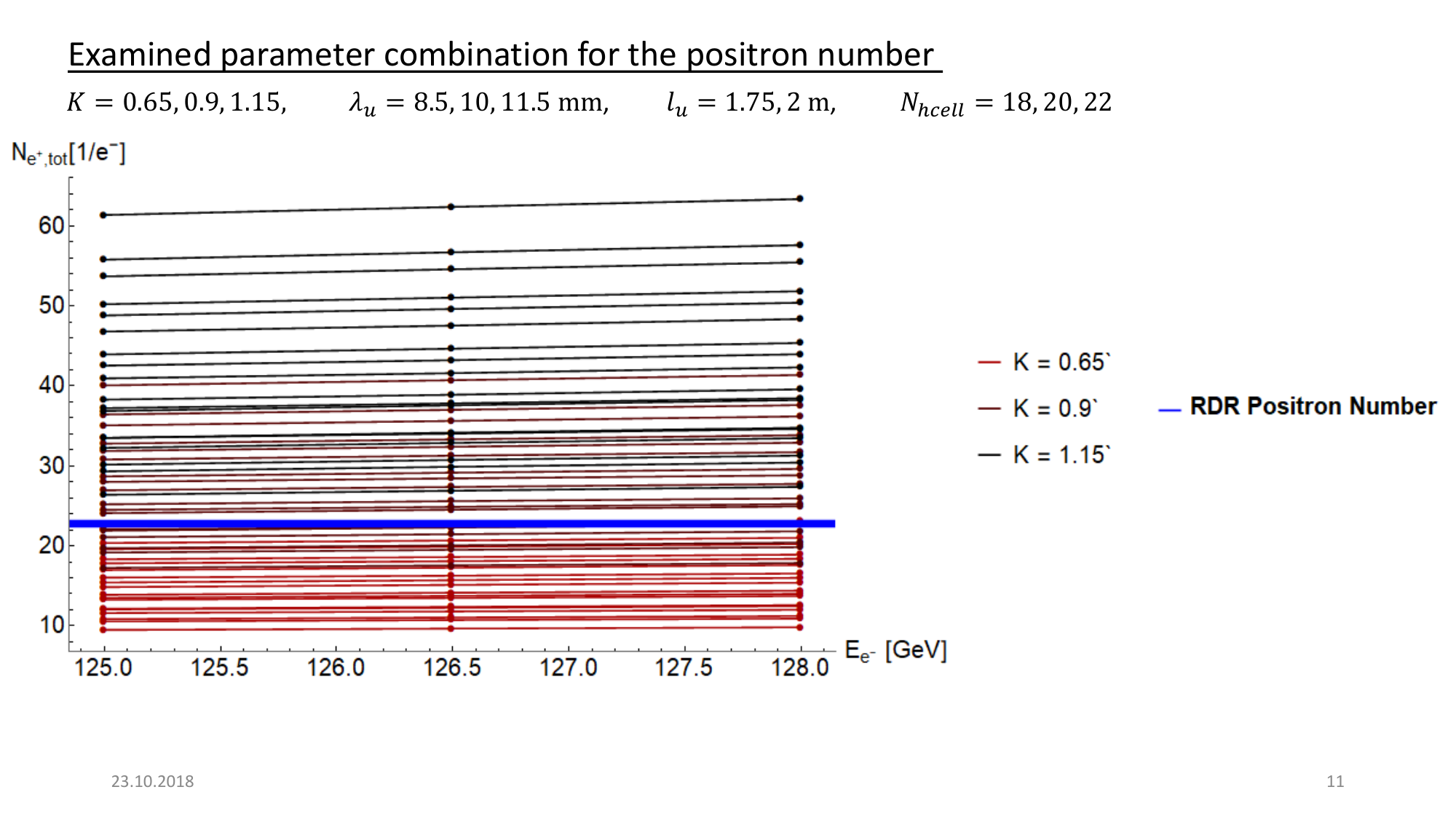}}
\caption{Resulting total number of positrons per electron for the different parameter variations but for the ideal case where all photons in the whole solid angle would hit the target. 
The blue line marks the expected positron number in the RDR set and has been regarded
for our study as a lower threshold ~\cite{bachelor}. \label{fig_9}}
\end{figure}

After discussing the deposited power $P_{vessel}$ in detail, we study now the total number of produced positrons $\dot N_{e^+}$ and try to maximize this quantity by altering 
the undulator parameter $K=0.65,$ 0.9, 1.15, the undulator period $\lambda_u=8.5,$ 10.0, 11.5~mm,  the undulator piece length $I_U=1.75$, 2~m 
and the number of half-cells $N_{hcell}=18$, 20, 22. We calculate  the total number of positrons produced per electron plotted against the electron drive beam energy, see 
Fig.~\ref{fig_9}.

\noindent
Please note, that the given number of positrons is the number, that would be produced only in the ideal case when all photons, which are emitted into the whole solid angle ($4 \pi$) would hit the target, so that the shown values in Fig.~\ref{fig_9} can only be taken as an upper limit. 

The number of lines shown in the graph is $54$, where each curve was calculated under a unique combination of values of the undulator parameter $K$, the undulator period 
$\lambda_u$, the undulator piece length $l_u$ and number of half-cells $N_{hcell}=18$ in the listed ranges.

The blue horizontal line at around $22$ positrons per electron denotes the result for the RDR parameters. 
This is regarded as our lower limit for the produced positron number and all curves below this line are dismissed in our study.
Obviously, there are still large gain factors possible even within the small variations we made in the parameters set-ups.

In Fig.~\ref{fig_{10}} we list all made parameter changes, ordered from highest to lowest number of resulting positrons. Those combinations that did not fulfill the 
threshold of RDR positron number are left out.
\begin{figure}[H]
	\fbox{
		\includegraphics[scale=0.5,page=12]{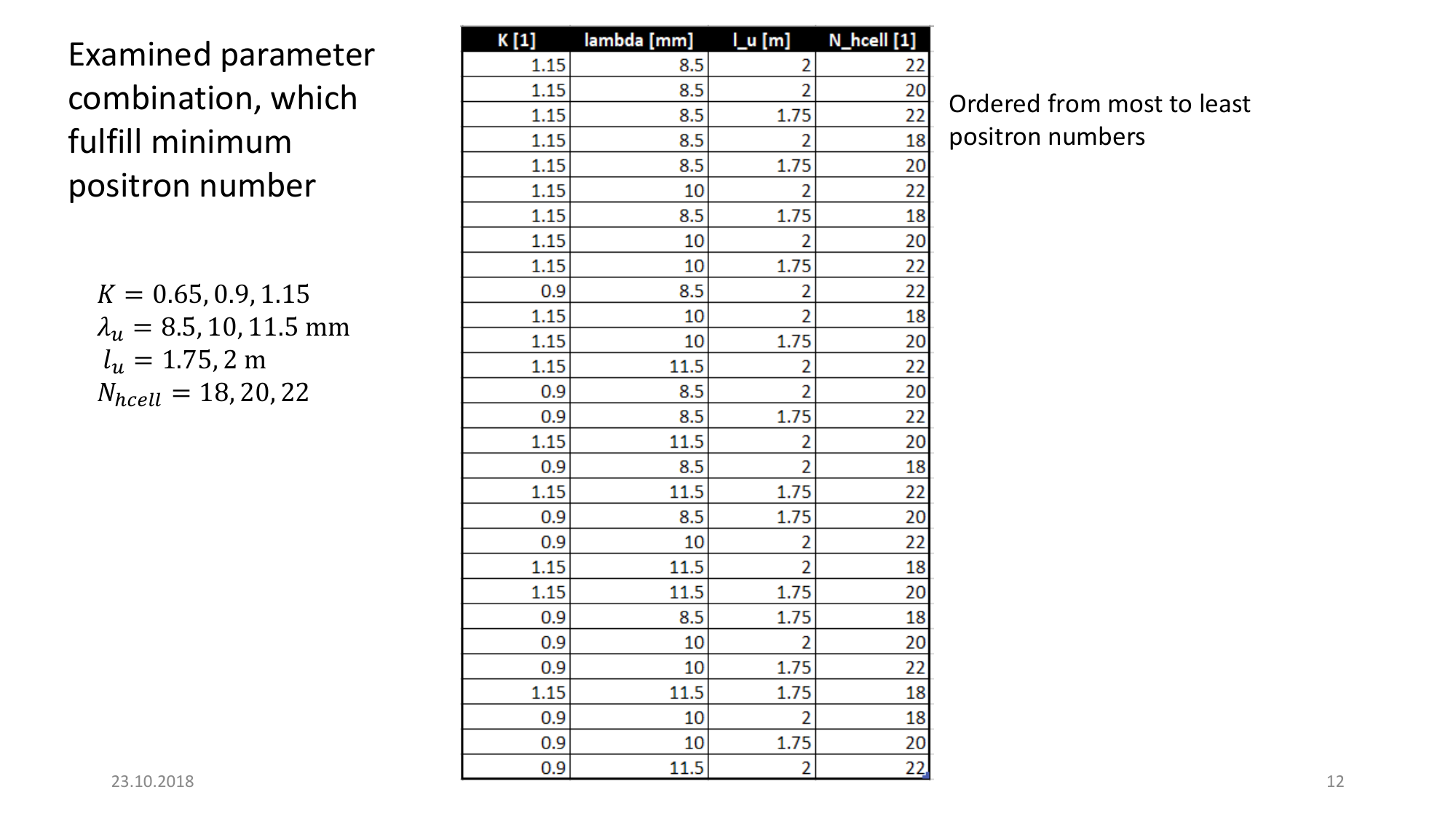}}
\caption{List of parameter combinations ordered from top to bottom
from the largest to the least number of produced positrons~\cite{bachelor}. \label{fig_{10}}}
\end{figure}
\noindent

\subsection{Test of the deposited power for parameter sets with highest $N_{e^+}$}
After deriving parameters sets that result in the highest positron number $N_{e^+}$, this set up has to be checked again with regard to the 
corresponding deposited power in the undulator 
vessel $P_{vessel}$.
\begin{figure}[H]
	\fbox{
		\includegraphics[scale=0.5,page=13]{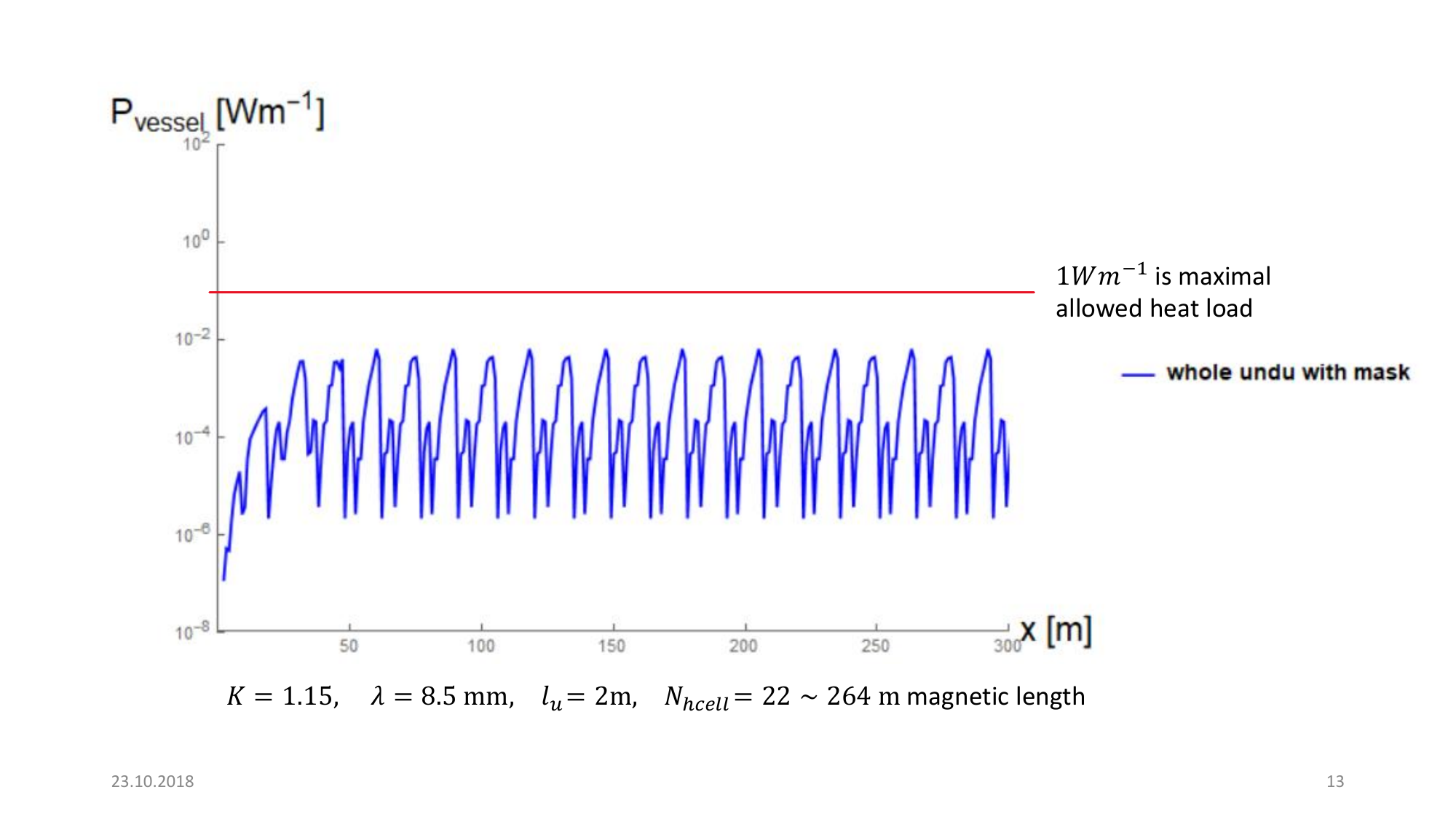}}
\caption{Resulting deposited power in the vessel after including the mask scheme for the parameter combination leading to the highest number of 
positrons, cf. also Fig.\ref{fig_{10}},~\cite{bachelor}. \label{fig_{11}}}
\end{figure}
\noindent
It can be seen from Fig.~\ref{fig_{11}} that the power limit of $\SI{1}{\W \per \m}$ is not exceeded even for the parameter set resulting in the 
largest number of positrons within the listed variations of the parameters and at $\sqrt{s}=250$~GeV.

\subsection{Further improvements}

\begin{figure}[H]
	\fbox{
		\includegraphics[scale=0.5,page=14]{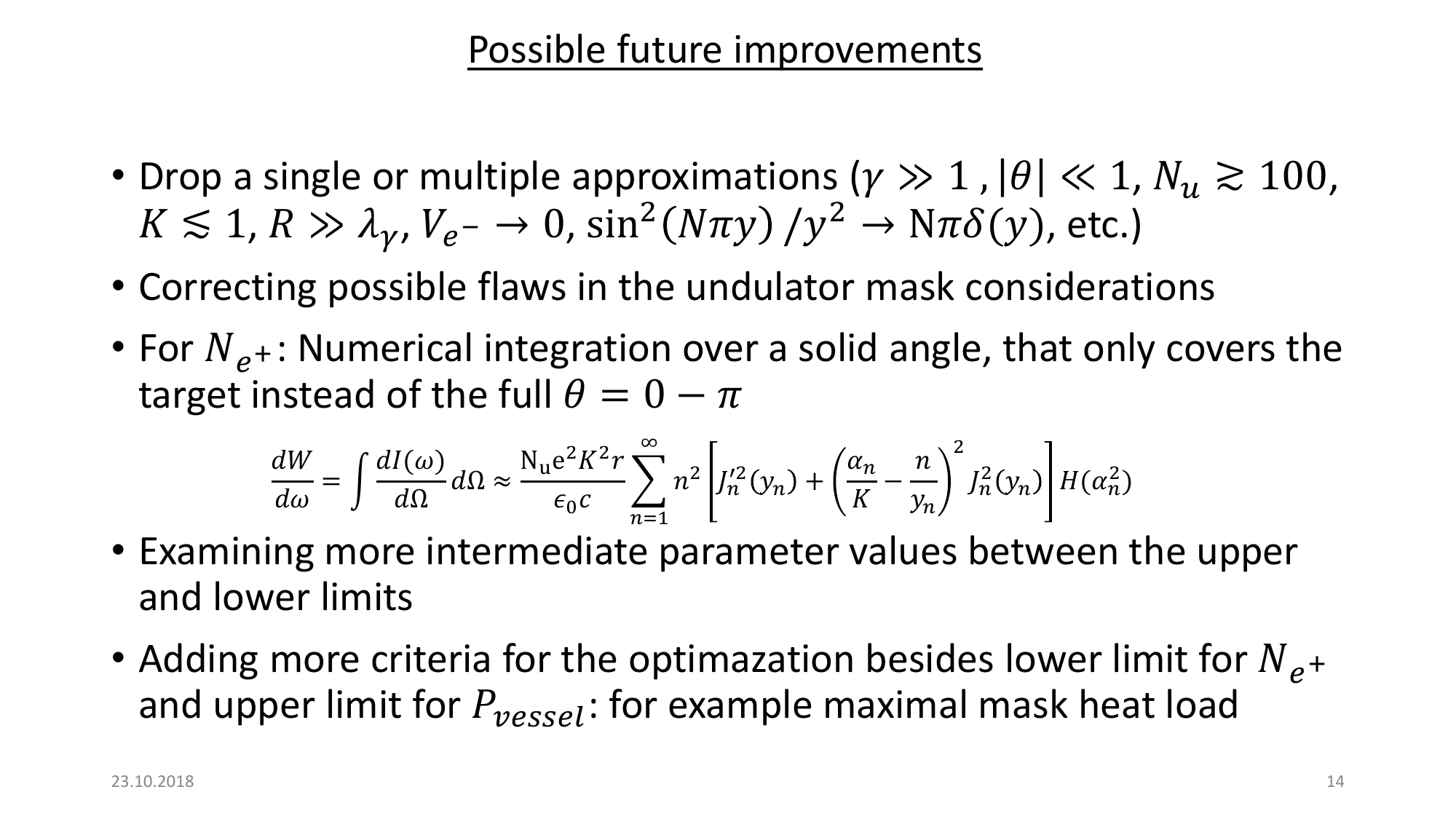}}
\caption{Next steps to improve the current study and towards further optimization of the positron yield~\cite{bachelor}.\label{fig_{12}}}
\end{figure}
The impact of several assumptions,  listed in the following, that have been made will be studied and removed.
For instance, we will drop systematically the made approximations, see item 1 in Fig.\ref{fig_{12}}, and study  their impact on the derived positron number.
Furthermore we include a fixed solid angle when integrating the number of positrons in order to cover only the target instead of the full angle.
Still possible flaws in the undulator mask will be studied in order to resolve the small deviations between ~\cite{duncan-thesis} and our current study~\cite{bachelor}.

Furthermore, we plan to perform the parameter scans on a more dense grid including engineering aspects and to extend the scanned
parameters, for example, to include also the maximal mask heat load, in order to optimize the number of positrons for the given set-up 
in even more detail and more precisely. These studies are currently under work~\cite{master}.

\section{Conclusions}
In this study we concentrated on finding optimized parameter sets for the given RDR undulator-scheme, originally 
designed for $\sqrt{s}=500$~GeV, for an cms-energy of only 250~GeV.
Since this energy step is physically of high interest for measuring the Higgs-boson couplings, mass and cross section, high luminosity is important already at this stage.
The availability of polarized beams is substantial in this regard. In order to match the promised high-precision measurements, however, a polarized positron beam is mandatory as well, otherwise several systematic effects can not be controlled~\cite{Robert-Thesis,Karl:2017xra,updateTDR}. Therefore it is very important to optimize the foreseen undulator parameter set via moderate changes. As our study has shown, a rather large positron gain factor  can be achieved even if only the magnetic field in the $K$ factor is slightly increased, the undulator period $\lambda_u$ 
is slightly decreased and the number of cells is slightly enhanced. All the made changes do not result in exceeding the limit of deposited power in the vessel.
Such promising results motivate the inclusion of further improvements concerning positron yield optimization for $\sqrt{s}=250$~GeV and  is currently under work.

\section*{Acknowledgements}
This work was supported by the German Federal Ministry of Education and Research,
Joint Research Project R\&D Accelerator ``Positron Sources'', Contract Number 05
H15GURBA.

\end{document}